\documentclass[
 aps, pra,
 amsmath,amssymb,
 11pt,
 final,
 tightenlines,
 twoside,
 twocolumn,
 nofloats,
 nofootinbib,
 superscriptaddress,
 showkeys,
 showkeywords,
]{revtex4-2}

\usepackage[T2A]{fontenc}
\usepackage[utf8]{inputenc}
\usepackage[russian,english]{babel}
\usepackage{graphicx}
\usepackage{dcolumn}
\usepackage{bm}
\usepackage{multirow} 

\def\squareforqed{\hbox{\rlap{$\sqcap$}$\sqcup$}}

\def\sq{\ifmmode\squareforqed\else{\unskip\nobreak\hfil
\penalty50\hskip1em\null\nobreak\hfil\squareforqed
\parfillskip=0pt\finalhyphendemerits=0\endgraf}\fi}

\def\utw{\smash{\rlap{\lower5pt\hbox{$\sim$}}}}

\def\udtw{\smash{\rlap{\lower6pt\hbox{$\approx$}}}}

\def\fm{\hbox{$\,.\!\!^{\rm m}$}}

\def\fdg{\hbox{$\,.\!\!^\circ$}}

\def\diameter{{\ifmmode\mathchoice
{\ooalign{\hfil\hbox{$\displaystyle/$}\hfil\crcr
{\hbox{$\displaystyle\mathchar"20D$}}}}
{\ooalign{\hfil\hbox{$\textstyle/$}\hfil\crcr
{\hbox{$\textstyle\mathchar"20D$}}}}
{\ooalign{\hfil\hbox{$\scriptstyle/$}\hfil\crcr
{\hbox{$\scriptstyle\mathchar"20D$}}}}
{\ooalign{\hfil\hbox{$\scriptscriptstyle/$}\hfil\crcr
{\hbox{$\scriptscriptstyle\mathchar"20D$}}}}
\else{\ooalign{\hfil/\hfil\crcr\mathhexbox20D}}%
\fi}}

\setcitestyle{authoryear,aysep={,},yysep={;},semicolon,round}

\makeatletter
\renewcommand \thefigure{\@arabic\c@figure} 
\renewcommand \thetable{\@arabic\c@table} 

\usepackage{xurl} 
\usepackage[breaklinks]{hyperref}

\hypersetup{
    colorlinks=true,
    linkcolor=blue,
    filecolor=magenta,
    urlcolor=blue,
    citecolor=blue,
    breaklinks=true 
}

\begin{document}

\selectlanguage{english}


\title{The massive binary system WR\,20a: light curve analysis in a colliding wind model}

\author{\firstname{I.~I.}~\surname{Antokhin}}
\email{igor@sai.msu.ru}
\affiliation{Sternberg Astronomical Institute, Moscow State University, Moscow, 119234 Russia}

\author{\firstname{E.~A.}~\surname{Antokhina}}
\affiliation{Sternberg Astronomical Institute, Moscow State University, Moscow, 119234 Russia}

\author{\firstname{A.~M.}~\surname{Cherepashchuk}}
\affiliation{Sternberg Astronomical Institute, Moscow State University, Moscow, 119234 Russia}

\begin{abstract}

The article presents the results of the analysis of optical light curves of the massive binary system WR\,20a (WN\,6ha\,+\,WN\,6ha). The analysis was performed with the binary system model, extending the standard Roche model for the case when both components of the system have powerful stellar winds. The model takes into account the collision of the winds and the influence of orbital motion on the collision zone. The observational light curves in the $BVI$ filters were taken from previously published papers, in which they were analyzed using the standard Roche model. The main difference between the results of our work and the previous results is that in our model the radii of the components are about 25\% smaller. As a consequence, the luminosity of the system in our model decreased by approximately 40\%, and the distance to the system by 20\%. In addition, the model was able to successfully describe the observed asymmetry of the light curve with respect to the phases of the conjunctions, which is impossible in the standard Roche model. The model light curves were also compared with the observational curves obtained by the TESS satellite and the ASAS-SN project. It was shown that, taking into account recent studies of interstellar extinction in the direction of the young open cluster Westerlund~2, the distance to WR\,20a obtained in our calculations is consistent with the hypothesis that WR\,20a is a member of the cluster.

\end{abstract}

\maketitle

\section{Introduction}\label{sec:intro}

The theory of stellar formation and structure cannot unambiguously predict the upper limit of stellar masses \citep[see, e.g.,][]{ulmer98}. Observational estimates of single star masses from spectrophotometric measurements depend on a variety of model assumptions, the assumed distances to clusters, and other factors. An example of the differences obtained in various studies is the most massive known star, RMC\,136a1. Its initial mass estimate $325\,M_\odot$ \citep{crowther16} was then changed to \mbox{$250$--$320\,M_\odot$} \citep{best20} and subsequently reduced to \mbox{$196\,M_\odot$} \citep{kalari22}. However, in a recent study by \cite{kes25} based on evolutionary calculations for single massive stars, a new value \mbox{$\sim 291\,M_\odot$} was obtained.

Perhaps one of the most objective methods for estimating the upper mass limit is to study the statistics of massive stars in relatively young clusters, such as Arches in our Galaxy \citep{figer05} or RMC\,136 in the Large Magellanic Cloud \citep{kalari22}. These studies show that the maximum stellar masses apparently do not exceed \mbox{$150$--$200\,M_\odot$}. However, it should be kept in mind that the absence of more massive stars in the observed samples may be due not only to a hypothetical fundamental limit on the maximum mass, but also to insufficient representativeness of the sample.

The most reliable estimates of stellar masses can be obtained from an analysis of radial velocity and light curves of eclipsing binaries. Since stars typically lose mass during their evolution (except in cases of mass exchange in close binary systems), the most massive stars should be expected in binary systems containing the youngest stars. For a long time, these were believed to be main sequence (MS) stars of spectral type O. For example, in our work \citep{ant2000}, we analyzed the light curve and determined the parameters of the components of the binary system HD\,93205 \mbox{(O3\,V((f))\,+\,O8\,V)}, containing an O3 component of the earliest spectral type known at that time.

However, in recent years it has become clear that the most massive stars are not those of O-type, but very luminous stars belonging to the Wolf-Rayet (WR) subtype~--- \mbox{WN\,5-7h} or \mbox{WN\,5-7ha}~--- nitrogen-sequence WR stars with signatures of hydrogen lines in their spectra. Such hydrogen-enriched WN stars have comparatively large radii for their masses and differ sharply from ``classical'' WR stars, which are in the late stages of evolution. They are comparatively young stars that have not yet descended (or have recently descended) from the MS, with hydrogen burning in the core, but with helium-enriched surface layers. The reason for the helium enrichment of the outer layers of these stars is not yet fully understood and is actively discussed \citep[see, e.g.,][]{egg06, schnurr09}. Hypotheses are being considered that the anomalous chemical composition of these stars is associated either with significant mass loss in the form of wind \citep{tutukov08} or with the mixing of matter in their interiors, stimulated by rapid axial rotation \citep{rauw05}.

WR\,20a is one of the most massive eclipsing binaries of this type known, consisting of two nearly identical stars of spectral type WN\,6ha\,+\,WN\,6ha. \cite{rauw04} conducted spectroscopic observations of the system and found that its components are WN\,6ha or O3\,If*/WN\,6ha stars. The orbit of the system is circular; the mass ratio inferred from radial velocity curves is \mbox{$M_2/M_1 = 0.99$}, and the minimum component masses are $M_1=70.7\pm 4.0\,M_\odot$ and \mbox{$M_2=68.8\pm 3.8\,M_\odot$}. \cite{bonanos04} observed WR\,20a in the $I$ band and found that the system is eclipsing. They analyzed the light curve using the Wilson-Devinney method in the standard Roche model for stars with thin atmospheres. In this model, the orbital inclination was found to be $i=74\,.\!\!^\circ5\pm 2\,.\!\!^\circ0$. At this inclination, the masses of the components were $M_1=83.0\pm5.0\,M_\odot$ and $M_2=82.0\pm5.0\,M_\odot$.

\cite{rauw05} obtained additional spectral observations of the system and refined the spectral types of the components: WN\,6ha\,+\,WN\,6ha. Based on non-LTE analysis of the spectra, the effective temperatures of the components were estimated, \mbox{$T_{1,2}=43\,000\pm 2000$\,K}, as well as the mass-loss rates, \mbox{$\dot{M}_{1,2} = 8.5 \times 10^{-6}\,M_\odot$\,yr$^{-1}$} (under the assumption that the component winds are non-uniform with the volume filling factor $f = 0.1$). \cite{rauw07} additionally obtained $BV$ light curves of the system and analyzed them together with the $I$ light curve \citep{bonanos04} in the standard Roche model for stars with thin atmospheres (the program {\tt NIGHFALL} was used, for details see \citealp{rauw07}). The orbital inclination $i=74\fdg 5\pm 1\fdg 0$ they found was the same as the value obtained by \cite{bonanos04}. While searching for the optimal solution, the effective temperature of the first (more massive) component was fixed at $T_1=43\,000$\,K (from the analysis of the spectrum in the non-LTE model), the optimal temperature of the second component was found to be \mbox{$T_2=40\,500$}\,K. The Roche lobe filling factors of both components were assumed to be equal, and their optimal value from the analysis of the light curves was found to be \mbox{$\mu_1=\mu_2=0.91$}, which corresponds to the stellar radii \mbox{$R_1\simeq R_2= 18.7\pm 0.9\, R_\odot$}. The values of the component radii and masses clearly indicated that both components are not very evolved stars, with hydrogen still burning in their cores (typical radii of the helium core in classical WR stars are \mbox{2--3\,$R_\odot$}).

Based on estimates of the bolometric luminosity and reddening of WR\,20a, \cite{rauw07} concluded that the system is a member of the young open cluster Westerlund\,2 (although the authors cautioned that various effects may change the distance estimates; we return to this issue in Section~\ref{sec:discussion}).

The main drawback of the described studies of WR\,20a light curves is that the standard Roche model used in \cite{bonanos04} and \cite{rauw07} is designed for stars with thin atmospheres. The presence of powerful stellar winds in both components of the system creates additional absorption. A model that does not take this into account overestimates the radii of the components. Moreover, since some model parameters correlate with each other, a change in the component radii can lead to a change in other model parameters. Therefore, in our work, we reanalyzed the WR\,20a light curves in our modified Roche model, which includes stellar winds of both components of the system \citep{ant24}.

Section~\ref{sec:obs} presents information on the observational light curves of WR\,20a used in our analysis. In Section~\ref{sec:results} the analysis of these curves and its results are described, Section~\ref{sec:discussion} provides a discussion of the obtained results, and Section~\ref{sec:concl} summarizes findings of the work.

\section{Observational light curves}\label{sec:obs}

Since one of the main goals of our study was a comparison with the results of \cite{bonanos04} and \cite{rauw07}, we used the observational $BV$ light curves from \cite{rauw07} and the $I$ light curve from \cite{bonanos04} to search for the model parameters. Note that the primary (more massive) component of the system at the orbital phase $\phi=0$ is located behind the secondary component. In the following, the primary component of the system will be denoted by the index ~1. The amplitude of the primary minimum of the $BVI$ light curves is about $0\fm 4$.

WR\,20a was also observed by the Transiting Exoplanet Survey Satellite (TESS). The TESS bandwidth covers the range 6000--10\,000\,\AA, the central wavelength is 7865\,\AA. The telescope observes a given sky sector measuring $24^\circ\times 96^\circ$ for about 27\,days at a time. An individual exposure is 2\,s, but the corresponding frames are summed up while on board the telescope, and the final exposure depends on the observing cycle. Access to public data is provided through the MAST\footnote{Barbara A. Mikulski Archive for Space Telescopes~--- \url{https://mast.stsci.edu/portal/Mashup/Clients/Mast/Portal.html}} database. In this database, light curves of the same object, based on the same original data, may be presented in several versions obtained by scientific teams participating in the TESS project. These versions differ in the processing method. For example, the SPOC (Science Processing Operations Center) and QLP (Quick Look Pipeline, \citealp{kunimoto22}) light curves were obtained using aperture photometry, while the PATHOS (A PSF-Based Approach to TESS High Quality Data of Stellar Clusters, \citealp{nar19}) light curves were obtained using PSF photometry for stars located in clusters. In addition, time intervals over which the original data are summed, differ.

\begin{figure}
\centering
\includegraphics[width=\columnwidth]{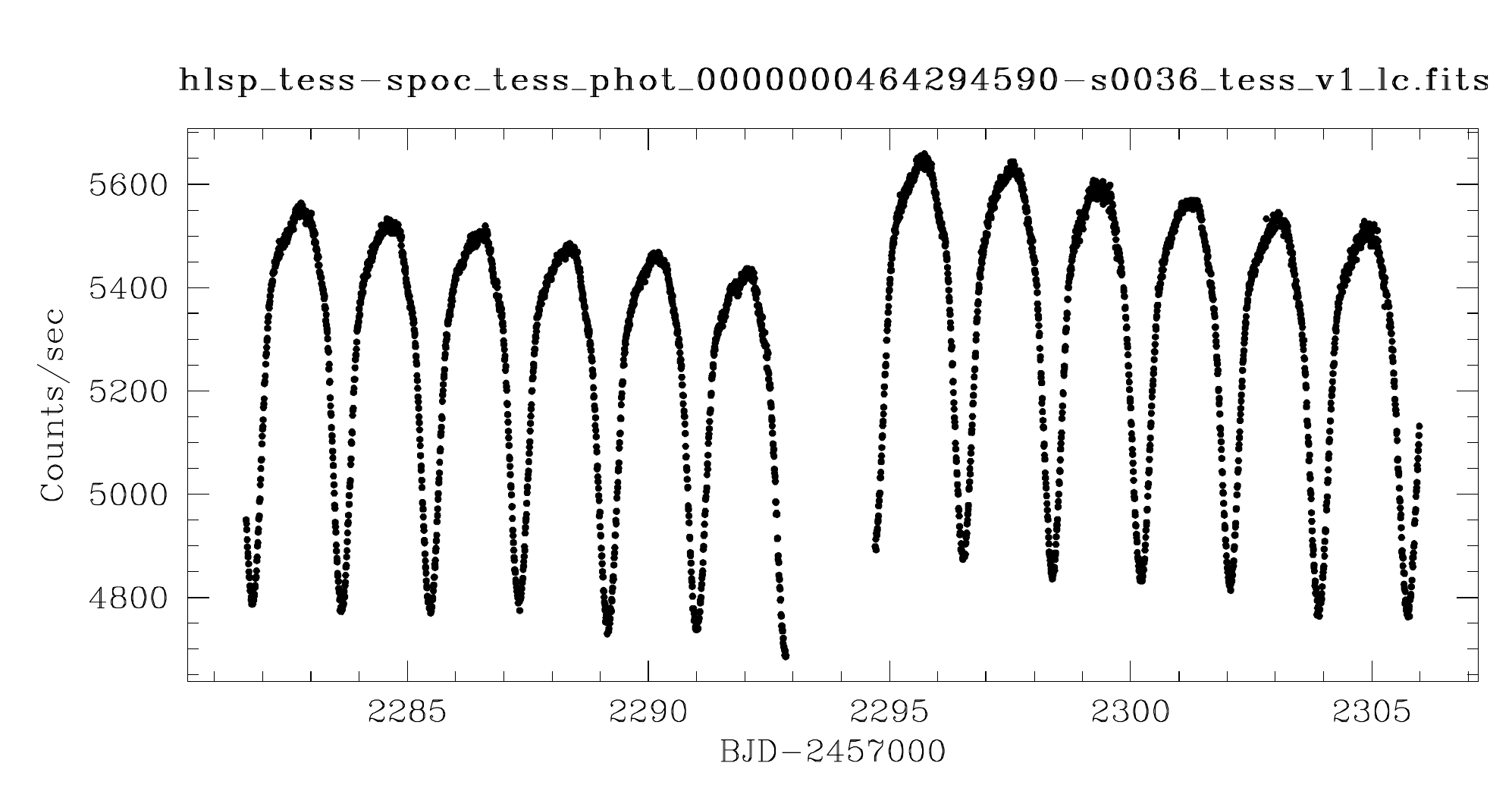}
\caption{An example of a light curve of WR\,20a obtained by the TESS satellite for sector 36 (cycle~3, \mbox{March--April} 2021) using simple aperture photometry (SAP) within the standard SPOC reduction. Systematic trends are clearly visible. Due to the influence of the ``third light'', the depth of the minima is about $0\fm 16$ (in the $BVI$ filters about  $0\fm4$).}
\label{TESS_lc_jd}
\end{figure}

We retrieved from the MAST database and compared all available light curves of WR\,20a, obtained both as a result of standard processing and by different teams. The pixel size of the TESS CCD is $15\,\mu$ (about $21^{\prime\prime}$), which corresponds to approximately the half-width of the PSF. Several other stars are located in the immediate vicinity of WR\,20a. As a result, the aperture photometry of WR\,20a is burdened by the presence of ``third light'' from nearby sources. The light curve amplitude for such photometry is noticeably smaller than the amplitudes of the light curves in \cite{bonanos04} and \cite{rauw07}. Moreover, the TESS light curves contain long-term trends. These effects are illustrated for one of the SPOC light curves in Fig.~\ref{TESS_lc_jd}. The figure shows the results of simple aperture photometry (SAP\_FLUX). The SPOC file for these observations also contains the light curve (column PDCSAP\_FLUX) to which the instrumental variability correction procedure has been applied. Unfortunately, in the case of WR\,20a, this procedure does not remove the trends present in the WR\,20a light curves and increases the amplitude of the ``corrected'' light curve to a wrong value, almost twice the amplitude of the light curve in the $BVI$ filters.

\renewcommand{\baselinestretch}{1.3}
\begin{table*}[!t]
\centering

\scriptsize

\caption{Files with TESS light curves for WR\,20a used in our work. Sector numbers, start and end dates of sector observations, and the number of brightness measurements are given. The different numbers of measurements for different reduction methods reflect the different intervals for summing individual exposures.}
\label{tab_tess_obs}

\begin{tabular}{l|c|c|c|r|c}
\noalign{\medskip}
 \hline
 \multicolumn{1}{c|}{File name} & Sector & Beginning    & End      & \multicolumn{1}{c|}{N} & Reduction \\
 \hline
 {\tt hlsp\_pathos\_tess\_lightcurve\_tic-0464294590-s0009\_tess\_v1\_ll.fits} & 9  & 28.02.2019 & 25.03.2019 &  1154 & PATHOS \\
 {\tt hls\_pathos\_tess\_lightcurve\_tic-0464294582-s0010\_tess\_v1\_llc.fits} & 10 & 26.03.2019 & 22.04.2019 &  1211 & PATHOS \\
 {\tt hlsp\_tess-spoc\_tess\_phot\_0000000464294590-s0036\_tess\_v1\_lc.fits}  & 36 & 07.03.2021 & 01.04.2021 &  3612 & SPOC   \\
 {\tt hlsp\_tess-spoc\_tess\_phot\_0000000464294590-s0037\_tess\_v1\_lc.fits}  & 37 & 02.04.2021 & 28.04.2021 &  3649 & SPOC   \\
 {\tt hlsp\_qlp\_tess\_ffi\_s0063-0000000464294590\_tess\_v01\_llc.fits}       & 63 & 10.03.2023 & 06.04.2023 & 11190 & QLP    \\
 {\tt hlsp\_qlp\_tess\_ffi\_s0064-0000000464294590\_tess\_v01\_llc.fits}       & 64 & 06.04.2023 & 04.05.2023 & 11356 & QLP    \\

 \hline
 \end{tabular}

\end{table*}
\renewcommand{\baselinestretch}{1.0}

The situation with PATHOS PSF photometry is significantly better, although problems with the light curve amplitude and trends were not completely eliminated. This is obviously due to the large pixel size relative to the PSF width. Unfortunately, not all available observations of WR\,20a have the PATHOS reduction option. Table~\ref{tab_tess_obs} lists the files from the MAST database that we used in this study. The total number of points on the light curves is 32172.

Given the presence of trends in most TESS data, as well as contamination by the ``third light'', we performed a correction for all of these data. The correction was based on two requirements:

\begin{list}{}{
\setlength\leftmargin{5mm} \setlength\topsep{0.5mm}
\setlength\parsep{0mm} \setlength\itemsep{0mm} }

    \item[(i)] the observed flux at the orbital phase 0.27 (the phase where the flux is maximum\footnote{The reasons why the flux is not maximum at the phase 0.25 will be discussed in Section~\ref{sec:results}.} should be constant;

    \item[(ii)] the amplitude of the primary minimum of the light curve should be equal to the amplitude of the primary minimum of the light curves in the $V$ and $I$ filters.

    \end{list}

After applying the correction, the TESS light curves were converted to a magnitude scale, and the magnitude at the maximum at the phase 0.27 was set to zero.

Photometric observations of WR\,20a are also available in the All-Sky Automated Survey for Supernovae (ASAS SN, \citealp{shappee14}; \citealp{hart23}) database. The project includes 24 ground-based telescopes around the world. We extracted the $V$-band light curve obtained in 2016–-2018, containing 192 measurements, from the project database. Since the pixel size of the CCDs in this project is also large, the data, as in the case of TESS, are contaminated by ``third light'' from nearby sources (the amplitude of the primary minimum of the light curve is about $0\fm17$). Trends are absent, so we applied only the ``third light'' correction to the ASAS-SN light curve and subsequent normalization of stellar magnitudes to the stellar magnitude at the phase 0.27.

\section{Analysis and results}\label{sec:results}

The model light curves were calculated using our new model of a binary system with colliding winds. The model is based on our previously proposed computer code for computing light and radial velocity curves of binary systems in the Roche model \citep{ant88,ant96,ant2000}. It is similar to the well-known code of Wilson and Devinney (\citealp{wilson71}; \citealp{wilson79}). As a development of this model, we earlier proposed a model of a binary system including a wind around one component \citep{ant13}. Using this model, we analyzed the light curve of the massive binary system WR\,22 (WN\,7h\,+\,O9\,V) (\citealp{craig22}; \citealp{ant23}). We then proposed another model \citep{ant24} that includes supersonic winds from both components, allowing it to compute light curves of systems in which both components possess powerful stellar winds. It was shown that accounting for absorption in the stellar winds leads to an increase in the width and depth of the minima compared to the standard Roche model. This model is used in the present paper. Its detailed description is given in \cite{ant24}; here we briefly summarize its main features.

The contact surface between the winds, as well as the densities of the cooling layers on either side of it, are calculated using a steady-state wind-wind collision model \citep{ant2004} in the radiative approximation. The collision zone's regime (radiative or adiabatic) is determined by the parameter $\chi$, which is the ratio of the characteristic gas cooling time to the characteristic gas outflow time from the collision region near the system axis \citep{stevens92}. The threshold value of the parameter $\chi\sim 1$. For WR\,20a $\chi\sim 0.02$, meaning the wind-wind collision occurs in the radiative regime. The velocity distribution of both winds is assumed to follow the generally accepted $\beta$-law \citep{cak75}. The set of model parameters includes both the standard parameters of the system and its components in the Roche model, and the parameters of the two winds: the mass-loss rates of each star $\dot{M}_1$, $\dot{M}_2$, the terminal wind velocities $V_{\infty,1}$, $V_{\infty,2}$, the parameters $\beta_{w,1}$, $\beta_{w,2}$ of the wind-velocity law, and the electron molecular weights of the winds matter $\mu_{e,1}$, $\mu_{e,2}$. The full list of the model parameters is given in Table~\ref{tab_model_params}.

\renewcommand{\baselinestretch}{1.3}
\begin{table}[!t]
\scriptsize
\caption{Model parameters. Indexes 1 and 2 refer to the primary and secondary components of the system.\medskip}
\label{tab_model_params}
\begin{tabular}{l|l} \hline
\multicolumn{1}{c|}{Parameter} &
\multicolumn{1}{c}{Description} \\
\hline
\multicolumn{2}{c}{Parameters of the Roche model} \\
\hline
$M_1\sin^3i$, $M_2\sin^3i$ & The masses of the components \\
                           & multiplied by the cube of the sine \\
                           & of the orbital inclination\\
$T_1$, $T_2$               & Mean effective stellar temperatures \\
$\mu_1$, $\mu_2$           & Roche lobe filling factors (the ratio \\
                           & of the polar radii of stars to the \\
                           & polar radii of the critical Roche lobe) \\
$i$                        & Orbital inclination angle \\
$P$                        & Orbital period \\
$e$                        & Eccentricity \\
$\omega_1$                 & Longitude of periastron for component 1 \\
$F_1,F_2$                  & Coefficients of asynchronous rotation \\
$\beta_1, \beta_2$         & Gravity-darkening coefficients \\
$A_1$, $A_2$               & Bolometric albedo \\
$x_1$, $y_1$, $x_2$, $y_2$ & Nonlinear limb-darkening coefficients \\
\hline
\multicolumn{2}{c}{Parameters of the winds} \\
\hline
$\dot{M}_1$, $\dot{M}_2$       & Mass-loss rates of the components \\
$V_{\infty,1}$, $V_{\infty,2}$ & Terminal velocities of the winds \\
$\beta_{w,1}$, $\beta_{w,2}$   & Parameters of the $\beta$-law  \\
$\mu_{e,1}$, $\mu_{e,2}$       & Electron molecular weights of the winds \\
\hline
\end{tabular}
\end{table}
\renewcommand{\baselinestretch}{1.0}

The main factor responsible for wind opacity in the optical continuum is electron scattering. Accordingly, the optical depth of each wind is proportional to

\begin{equation}
 \tau_0 = \sigma_T n_0 a\,,
\end{equation}

where $a$ is the size of the semi-major axis of the system orbit, $n_0$ is the reference electron density (the wind density at a distance $a$ from a component, assuming that the wind speed at this point is $V_\infty$)

\begin{equation}\label{eq_n0}
 n_0 = \cfrac{\dot{M}}{4\pi m_p\mu_ea^2V_\infty}\,.
\end{equation}

Here $m_p$ is the proton mass, and $\mu_e$ is the electron molecular weight. Obviously, $\dot{M}$, $V_\infty$ and $\mu_e$ are not independent parameters of our model, therefore, it is generally convenient to use $\tau_0$ as a free model parameter. However, in our study, we used fixed values of $V_\infty$ and $\mu_e$ for both components of the system (see below). Therefore, we used $\dot{M}_1$ and $\dot{M}_2$ as free model parameters.

Some model parameters can be fixed. The values of $P$, $M_1\sin^3i$, $M_2\sin^3i$, $e$ were determined from spectroscopic observations of \cite{rauw05}. The orbit of the system is circular, therefore $e=0$, $\omega_1=0\,\deg$. The analysis of light curves allows us to determine only the ratio of the component temperatures. Therefore, as in \cite{rauw07}, we fixed the temperature of the primary component: $T_1=43\,000$\,K. The rotation of the components was assumed to be synchronous with the orbital motion, thus the corresponding coefficients $F_1$, $F_2$ were set equal to unity. The gravity-darkening coefficients and albedos were chosen according to theoretical values for radiative envelopes from \cite{zeip24} and \cite{ruc69}, respectively. The values of the limb-darkening coefficients (square-root law) were taken from \cite{hamme93}.

Precise determination of the chemical composition of the WR\,20a components is a challenging task. According to \cite{rauw05}, the abundance of helium is much lower than that of hydrogen, the nitrogen abundance is six times solar, and the carbon abundance is ``much lower'' than that of the Sun. Because of the uncertainty in the chemical composition, we fixed the $\mu_e$ values for both winds to match the solar chemical composition. If necessary, the obtained $\dot{M}_1$, $\dot{M}_2$ values can be recalculated taking into account the new $\mu_e$ values using the formula \eqref{eq_n0}. The $V_\infty$ and $\beta_w$ values for the winds of both components were the same as in \cite{rauw05}.

Thus, the model has six free parameters:

\begin{list}{}{
\setlength\leftmargin{5mm} \setlength\topsep{0.0mm}
\setlength\parsep{0mm} \setlength\itemsep{0mm} }
    \item[$\bullet$]the orbital inclination angle $i$;

    \item[$\bullet$]the temperature of the secondary component $T_2$;

    \item[ $\bullet$]the Roche lobe filling factors $\mu_1$, $\mu_2$;

    \item[$\bullet$]the mass-loss rates $\dot{M}_1$, $\dot{M}_2$.

\end{list}

It is well known that even in the standard Roche model, some parameters are interrelated. In our model, an additional anticorrelation between stellar radii and mass-loss rates may be observed, since an increase in the latter plays the same role as an increase in stellar radii in the standard Roche model. Note, however, that this is only true for a fixed temperature ratio of the components. If it changes, the correlations become more complex. This means that optimizing the model for the six parameters listed above may lead to a formally satisfactory solution that turns out to be physically inconsistent. Therefore, in addition to the case where all six parameters were assumed to be free, we performed optimization for several other cases implementing various additional constraints.

The part of our code responsible for computing the light curves of binary systems in the standard Roche model and the {\tt NIGHTFALL} code used in \cite{rauw07} are independent implementations of the Wilson--Devinney code. Therefore, before starting calculations in our two-wind model, it was necessary to make sure that the model light curves obtained with these two codes in the case of the standard Roche model were the same. To verify this, we calculated the model light curves in the $BVI$ filters, ``switching off'' the component winds in our model and using the parameters of the Roche model from \cite{rauw05,rauw07}. As expected, our model light curves agree with the model light curves in Fig.~5 in \cite{rauw07} with very good accuracy.

Model optimization was performed by the Markov Chain Monte Carlo method (MCMC). This method uses a Bayesian approach and yields robust posterior probability distributions and estimates of model parameter errors. Critical to the iteration process is the optimal selection of the so-called proposed variance. We employ a variant of the method called ``Adaptive Metropolis within Gibbs'' \citep{mcmc11}, which adaptively and automatically selects the optimal variance without destroying the statistics of the posterior distributions. Details of our implementation are described in \cite{ant22}.

The number of MCMC iterations for each run was 100\,000. Since the MCMC variant we used changes only one parameter at each step, $10\,n$ steps ($n$ is the number of free variables) were taken between subsequent iterations to give each parameter a chance to change. Thus, for example, for the case with six free parameters, the direct problem of calculating the model light curve for a given set of parameters was solved $6.1\times 10^6$ times. The characteristic burn-in period of the Markov process was 2--3 thousand iterations.

Before presenting the results of the main simulation, we present the result of the model optimization, in which all parameters, except for the Roche lobe filling factors $\mu_1$ and $\mu_2$, were fixed at the values from \cite{rauw05,rauw07}, and $\mu_1$ and $\mu_2$ were assumed equal, as in the cited work. Thus, in this case there is only one free model parameter $\mu=\mu_1=\mu_2$. Solving the problem resulted in $\mu_1=\mu_2=0.773\pm 0.009$. This value is smaller than the value 0.91 from the work of \cite{rauw07} (the reason for this will be discussed below, in Section~\ref{sec:discussion}). Obviously, a change in the Roche lobe filling factors can entail a change in other parameters. Therefore, the main part of the calculations was carried out with several free parameters of the model.

As expected, in the case of six free model parameters, the solution proved physically inconsistent. The optimization algorithm was unable to select a unique set of optimal parameters and switched between two formal solutions:

\begin{list}{}{
\setlength\leftmargin{5mm} \setlength\topsep{0mm}
\setlength\parsep{0mm} \setlength\itemsep{0mm} }
    \item[(i)]$\mu_1\simeq 0.55$, $\mu_2\simeq 0.86$, \\ $i\simeq 77^\circ$, $T_2\simeq 39\,500$\,K, \\ $\dot{M}_1\!\simeq\! 0.62\!\times\!10^{-5}\,M_\odot$\,yr$^{-1}$, \\ $\dot{M}_2\!\simeq\!1.52\!\times\!10^{-5}\,M_\odot$\,yr$^{-1}$

    and

    \item[(ii)]$\mu_1\simeq 0.85$, $\mu_2\simeq 0.45$, \\ $i\simeq 77^\circ$, $T_2\simeq 58\,800$\,K, \\ $\dot{M}_1\!\simeq\! 1.44\!\times\!10^{-5}\,M_\odot$\,yr$^{-1}$, \\ $\dot{M}_2\!\simeq\!1.04\!\times\!10^{-5}\,M_\odot$\,yr$^{-1}$.
\end{list}

In these two solutions, the parameters of the components are essentially swapped. Furthermore, in each solution, the parameters of the components are significantly different. However, as follows from the spectral analysis, as well as from the fact that the depths of the primary and secondary minima in the light curves are virtually identical and do not vary with wavelength, the components of the system should have very similar characteristics.

In addition, we ran several optimizations that imposed additional constraints on the parameters, such as fixing $T_2=40\,500$\,K (as in \citealp{rauw07}), assuming that $\dot{M}_1=\dot{M}_2$ or $\mu_1=\mu_2$, and leaving the remaining parameters free, etc. In all these cases, we also obtained solutions in which the component parameters differed significantly. For the same reason, \cite{rauw07} optimized their model by assuming equal Roche lobe filling factors. After performing the described experiments, for the final model optimization, we chose the case in which $\mu_1=\mu_2$ and $\dot{M}_1=\dot{M}_2$. Thus, in this case there were four free model parameters: $i$, $T_2$, $\mu$ ($\mu_1=\mu_2$), \mbox{$\dot{M}$ ($\dot{M}_1=\dot{M}_2$)}.

\begin{figure}
\centering
\includegraphics[width=\columnwidth]{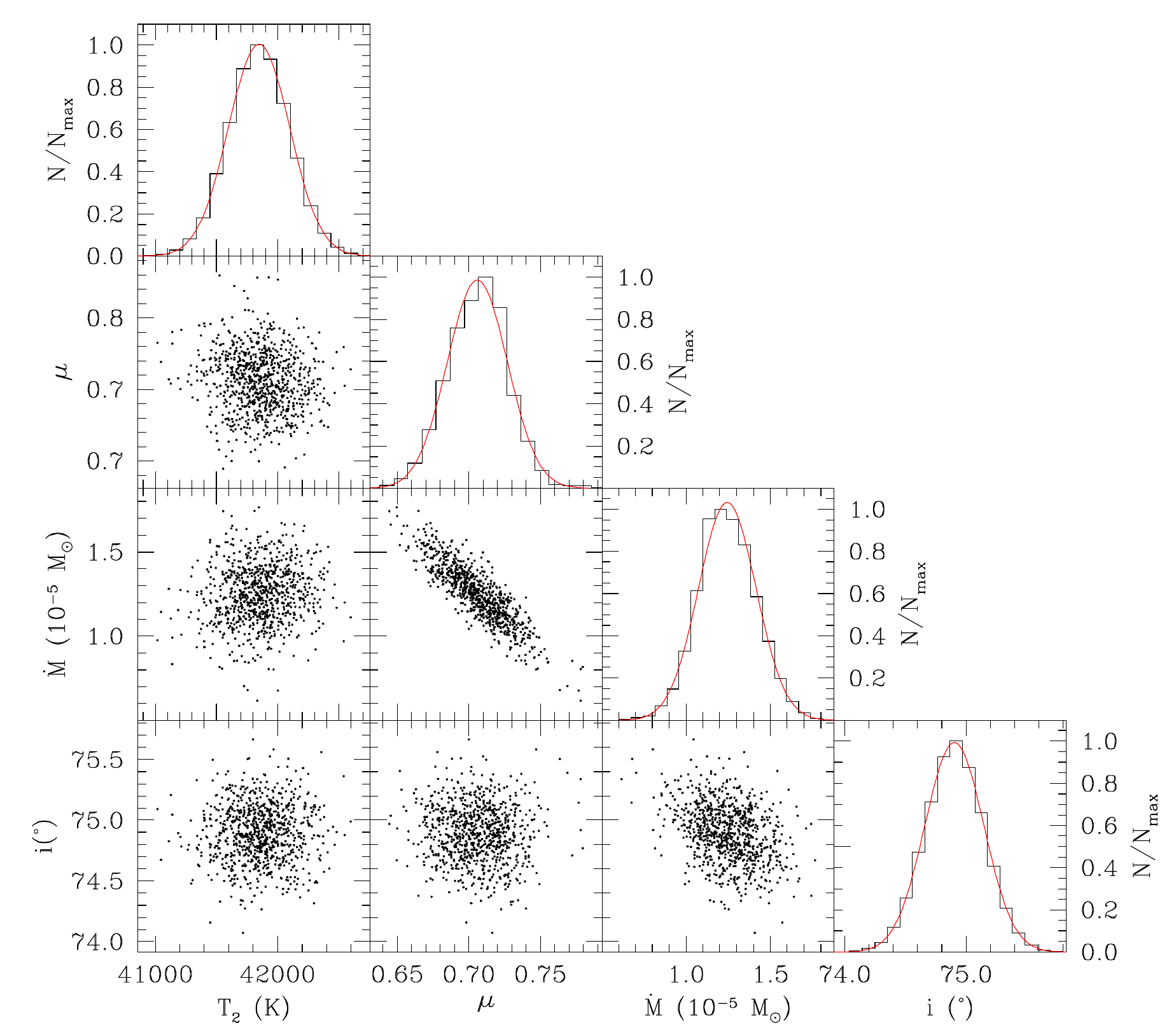}
\caption{Posterior empirical distributions for all possible pairs of parameters $i$, $T_2$, $\mu$, $\dot{M}$. To avoid cluttering the figure, only every hundredth point is shown. Shown on the right (top) of each row are the empirical histograms of the probability distribution for a given parameter (normalized to the maximum) and the Gaussian functions approximating them (shown by solid red curves).}
\label{fig_qq}
\end{figure}

\begin{figure}
\centering
\includegraphics[width=\columnwidth]{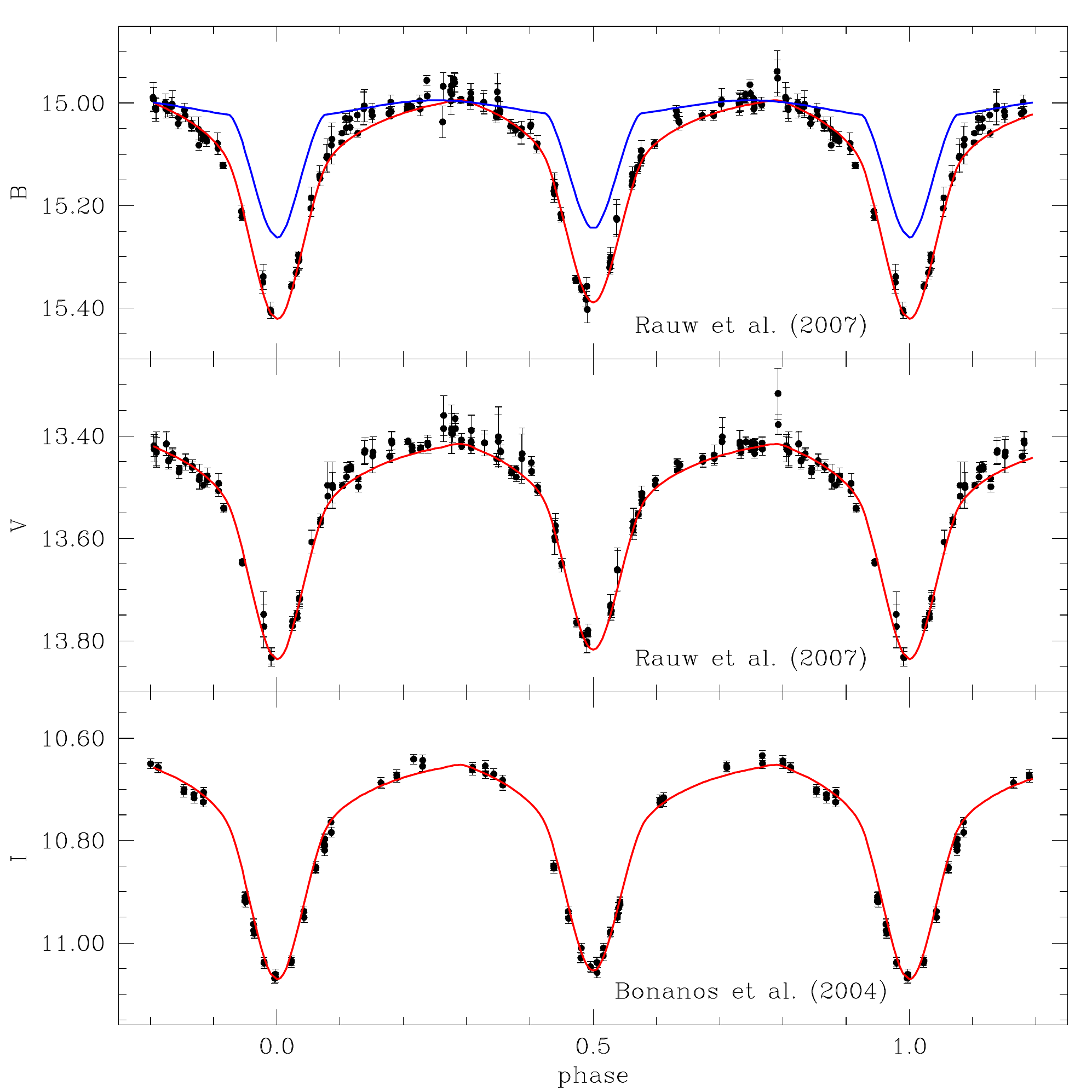}
\caption{Observational light curves in the $BVI$ filters from \cite{bonanos04}, \cite{rauw07} (dots) and the optimal model light curves in our model (solid red lines). Note that in \cite{bonanos04} the uncertainties for all observational points in the $I$ filter are the same ($0\fm 01$) and, obviously, are underestimated. For comparison, the model light curve for our optimal model, in which the component winds are ``switched off'' (see text), is shown in blue in the top panel (filter $B$).}
\label{fig_lc_BVI}
\end{figure}

The best fit model for the $BVI$ light curves from \cite{bonanos04} and \cite{rauw07} is shown in Figs.~\ref{fig_qq},~\ref{fig_lc_BVI} and in Table~\ref{tab_found_params}. The last column of this Table lists the parameters taken from \cite{rauw05,rauw07}. Note that the wind parameters in these papers were determined not from the light curve solution (the authors used the standard Roche model), but from the analysis of the WR\,20a spectrum.

Fig.~\ref{fig_qq} shows the empirical posterior distributions of the model parameters obtained by applying the MCMC method, as well as the corresponding histograms and the Gaussian functions approximating them for each parameter. The mean values and standard deviations calculated from the empirical distributions were chosen as the optimal model parameters and their errors. The anticorrelation of the Roche lobe filling factor $\mu$ with the mass-loss rate $\dot{M}$ is clearly visible, the latter also weakly anticorrelates with the orbital inclination angle $i$.

Figure~\ref{fig_lc_BVI} shows the observational light curves in the $BVI$ filters from \cite{bonanos04} and \cite{rauw07}, as well as the best-fit model light curves (in red). It is evident that the observational light curves are asymmetric with respect to the conjunction phases 0 and 0.5. This asymmetry is well reproduced by our model curves. In the model, it arises from the inclination of the contact surface to the system axis, caused by the Coriolis force. The inclination is calculated in the model using a theoretical formula, and is not specified as a model parameter. The standard Roche model cannot describe this asymmetry, since in this model the geometry of the system, which is in a circular orbit, is symmetric with respect to the conjunction phases. The presence of the wind-wind collision zone, inclined to the system axis, disrupts the axial symmetry of the system and makes the light curve asymmetric, with maxima at phases 0.27 and 0.77. For comparison, the top panel ($B$ filter) shows in blue the model light curve for the same system parameters, but with winds of both components ``switched off''. The difference between the two model light curves illustrates the contribution of the component winds to opacity.

However, neither the standard Roche model used in \cite{rauw07} nor our model can accurately describe the observational light curves (at least in the $BV$ filters) in the phase range \mbox{0.1--0.3}, where the observational light curves are higher than at the similar phases after the secondary minimum \mbox{0.6--0.8}. We will return to this issue below.

\renewcommand{\baselinestretch}{1.3}
\begin{table*}[t]
\caption{The values of the best fit model parameters.}
\label{tab_found_params}
\begin{tabular}{l|c|c|c}
\noalign{\medskip}
\hline
\multicolumn{1}{c|}{\multirow{2}{*}{Parameter}} & The value in  & \multicolumn{1}{c|}{\multirow{2}{*}{Status}} & The value from \\[-1mm]
                             & our model &                             & \cite{rauw05,rauw07} \\
\hline
\multicolumn{4}{c}{Roche model parameters}\\
\hline
$M_1\sin^3i$, $M_\odot$  &  $74.0$    & Fixed     & $74.0$  \\
$M_2\sin^3i$, $M_\odot$  &  $73.3$    & Fixed     & $73.3$  \\
$T_1$, K                 & $43\,000$    & Fixed     & $43\,000$ \\
$T_2$, K                 & $41\,840\pm 250$  & Free  & $40\,500$ \\
$\mu$                    & $0.71\pm 0.02$  & Free, $\mu_1=\mu_2$ & $0.91$  \\
$i$, deg                 & $74.9\pm 0.3$   & Free & $74.5\pm 1.0$ \\
$P$, days                & $3.68475$  & Fixed     & $3.68475$ \\
$e$                      & $0.0$      & Fixed     & $0.0$ \\
$\omega_1$, deg          & $0.0$      & Fixed     & $0.0$ \\
$F_1,F_2$                & $1.0$      & Fixed     & \\
$\beta_1, \beta_2$       & $0.25$     & Fixed     & \\
$A_1$, $A_2$             & $1.0$      & Fixed     & \\
\hline
\multicolumn{4}{c}{Wind parameters}\\
\hline
$\dot{M}$, $M_\odot$\,yr$^{-1}$    & $(1.24\pm 0.17)\times 10^{-5}$  & Free, $\dot{M}_1 = \dot{M}_2$ & $0.85\times 10^{-5}$ \\
$V_{\infty,1}$, km\,s$^{-1}$           & $2800$                  & Fixed   & $2800$ \\
$V_{\infty,2}$, km\,s$^{-1}$           & $2800$                  & Fixed   & $2800$ \\
$\beta_{w,1}$               & $1.0$                   & Fixed   & $1.0$  \\
$\beta_{w,2}$               & $1.0$                   & Fixed   & $1.0$  \\
$\mu_{e,1}$                 & $1.17$                  & Fixed   &        \\
$\mu_{e,2}$                 & $1.17$                  & Fixed   &        \\
\hline
\multicolumn{4}{p{12.5cm}}{\footnotesize
We do not provide numerical values for the limb-darkening coefficients $x$ and $y$, since for each filter, stellar temperature, and logarithm of gravity, they were selected from tabulated values according to \cite{hamme93}.

The values of some parameters in the last column are not given, since they were not provided explicitly in \cite{rauw05,rauw07}.}

\end{tabular}
\end{table*}
\renewcommand{\baselinestretch}{1.0}

\begin{figure}
\centering
\includegraphics[width=0.818\columnwidth]{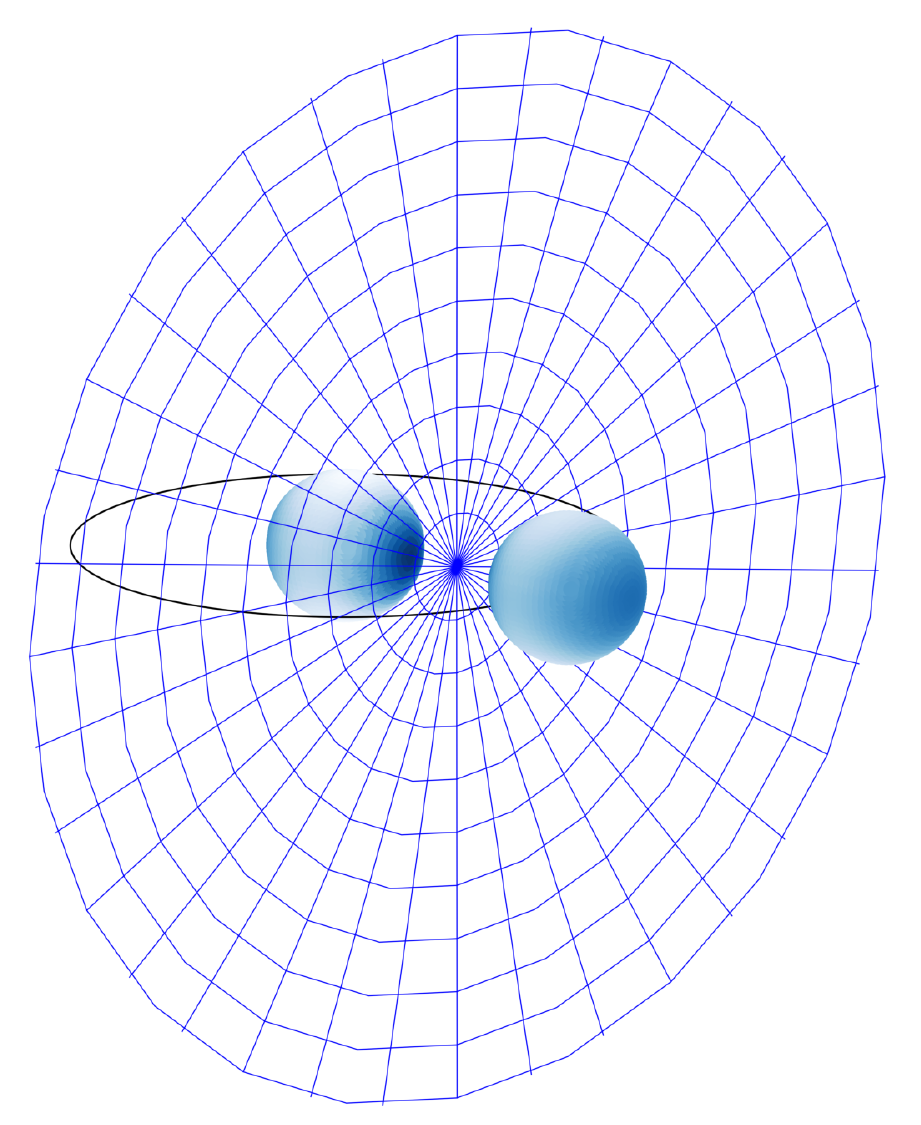}
\caption{The view of the system as seen by an observer at the orbital phase 0.15. The orbit is shown by the black line, with the primary component behind and the secondary in front. The contact surface is shown in blue. To avoid clutter, only the central part of the contact surface is shown; the mesh on it is much more sparse than in real calculations.}
\label{fig_cartoon}
\end{figure}

Fig.~\ref{fig_cartoon} shows a view of the system in our model, as seen by an observer at the orbital phase 0.15 (exit from the primary minimum).

\begin{figure}
\centering
\includegraphics[width=\columnwidth]{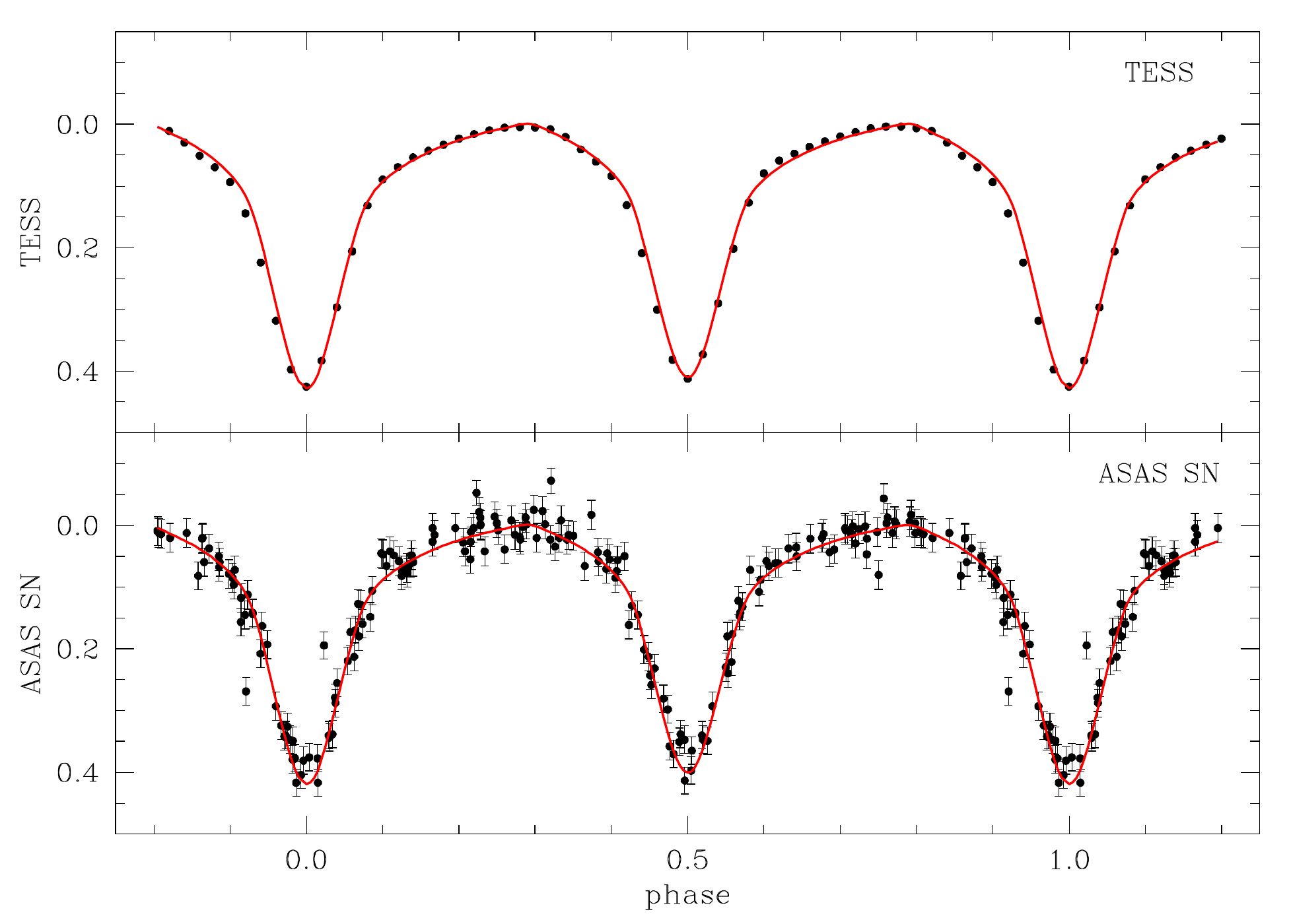}
\caption{Top panel: the mean observed TESS light curve (dots) and the model light curve (solid red line) in a model with parameters found from $BVI$ filters. The uncertainties in the mean observed light curve are comparable to the size of the dots. Bottom panel: The ASAS SN light curve and the similarly computed model light curve.}
\label{fig_lc_TESS_ASAS}
\end{figure}

Figure~\ref{fig_lc_TESS_ASAS} shows the observational and model light curves for the TESS and ASAS-SN data. The top panel shows the average TESS light curve; the phase bins were set to 0.02 of the orbital periods. Recall that the observational curves were corrected to remove trends and the influence of the ``third light''. The corrections were based on the characteristics of the observational light curves in the $VI$ filters, so the TESS and ASAS-SN data cannot be considered as an independent source of information. Therefore, we did not use them to find the optimal model parameters; instead, we calculated the model light curves for the corresponding passbands with the model parameters found from the $BVI$ data. Nevertheless, given the large number of measurements (especially in the TESS data), these data are of interest. First, they allow us to refine the asymmetry of the observational light curves with respect to the phases of conjunctions and to test how well the model light curves fit this asymmetry. Second, they allow us to further compare the relative stellar magnitudes of the observational light curves at phases 0.1--0.3 and 0.6--0.8. Third, they allow us to clarify the difference (if any) between the depths of the primary and secondary minima, which was not entirely clear from observations in the $BVI$ filters. Recall that in the process of removing trends from the TESS data, we did not require that the brightness at the two maxima of the light curve be equal. Only the brightness at the phase 0.27 was equalized. Similarly, the ``third light'' correction was performed so that only the depth of the primary minimum was kept equal to the depth of the primary minimum in the $VI$ filters; the depth of the secondary minimum was obtained automatically.

\begin{figure}
\centering
\includegraphics[width=\columnwidth]{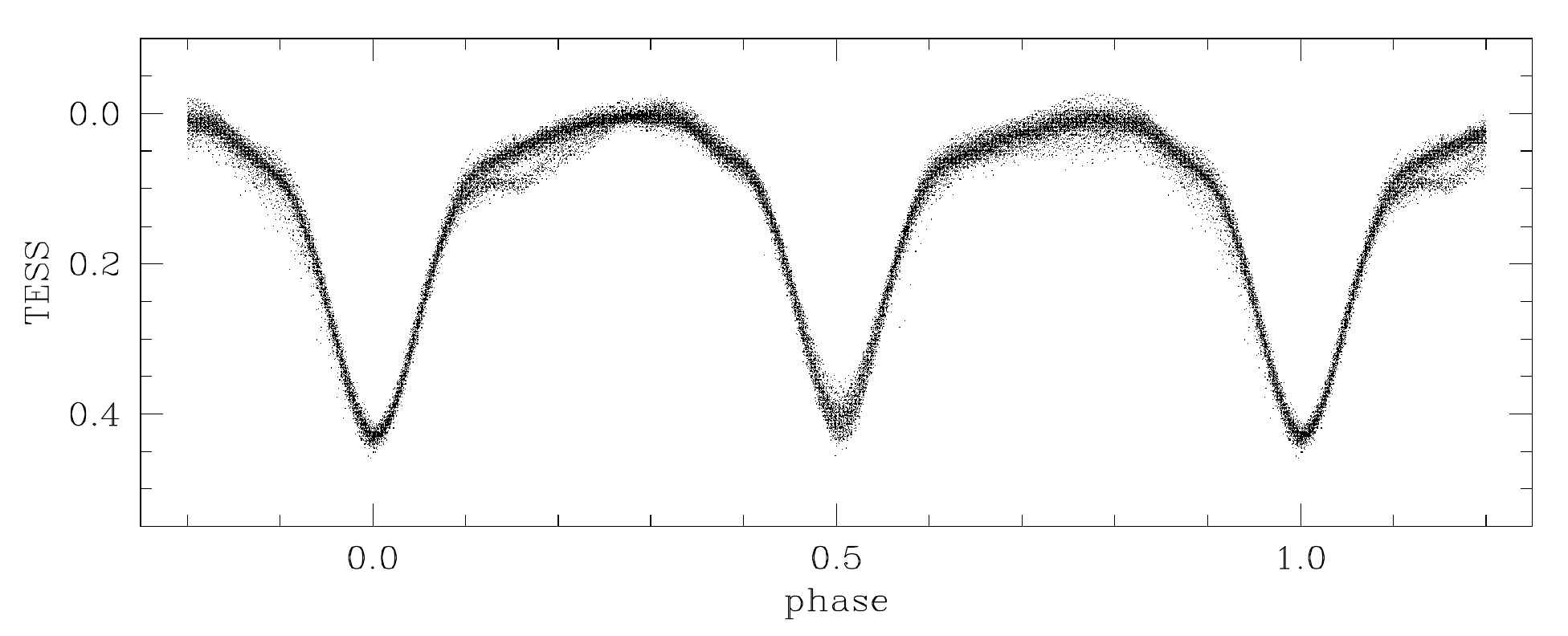}
\caption{TESS observational light curve containing individual measurements folded with the orbital period.}
\label{fig_lc_TESS_indiv}
\end{figure}

As can be seen from Fig.~\ref{fig_lc_TESS_ASAS}, there is a distinct asymmetry in the observational light curves with respect to the conjunction phases. It is well fit by our model. The depth of the secondary minimum is indeed slightly smaller than the depth of the primary, although the difference is very slight. At the same time, the excess of the brightness level at phases 0.1--0.3 over the level at phases 0.6--0.8, clearly observed in the $BV$ light curves (the data in the $I$ filter are too sparse to make a definitive judgment), is absent in the TESS and ASAS-SN data. The explanation for this difference is probably related to non-stationary processes. Indeed, such processes occur in the wind-wind collision zone, which can lead to variability on both short and long time scales. The presence of irregular variability in the light curve is illustrated by Fig.~\ref{fig_lc_TESS_indiv}, which shows the non-averaged TESS light curve. The $BV$ light curves were obtained over a period of one and a half months in late 2004--early 2005, while the ASAS-SN observations were carried out in 2016--2018, and TESS in 2019--2023. The natural suggestion is that the $BV$ light curves of \cite{rauw07} reflect the ``instantaneous'' state of non-stationary processes in the system. The mean TESS and ASAS-SN light curves reflect the average state of the collision zone. Our steady-state model, by its very nature, describes the time-averaged state of the collision zone and, for this reason, describes these light curves well.

\section{Discussion}\label{sec:discussion}

\renewcommand{\baselinestretch}{1.3}
\begin{table}[t]
\caption{The absolute parameters of the best-fit model.}
\label{tab_abs_params}
\begin{tabular}{l|r@{$\,\pm\,$}l|r@{$\,\pm\,$}l}
\noalign{\medskip}
\hline
\multicolumn{1}{c|}{\multirow{3}{*}{Parameter}}                    &\multicolumn{4}{c}{Value}  \\
\cline{2-5}
                         & \multicolumn{2}{c|}{\multirow{2}{*}{Our model} }   & \multicolumn{2}{c}{\citeauthor{rauw05}} \\
                         &  \multicolumn{2}{c|}{ }    & \multicolumn{2}{c}{ (\citeyear{rauw05,rauw07}) } \\
\hline
$M_1$, $M_\odot$        &   82.2 & 4.7       & 82.7& 4.8$^{a)}$ \\
$M_2$, $M_\odot$        &   81.4 & 4.7       & 81.9& 4.8$^{a)}$ \\
$R_1$, $R_\odot$        &   14.1 & 0.4$^{b)}$ & 18.7& 0.9  \\
$R_2$, $R_\odot$        &   14.0 & 0.4$^{b)}$ & 18.7& 0.9  \\
$\log L_1/L_\odot$      &   5.78 & 0.09      & 6.03& 0.09 \\
$\log L_2/L_\odot$      &   5.73 & 0.09      & 5.93& 0.10 \\
$\log (L_1+L_2)/L_\odot$&   6.06 & 0.09      & 6.28& 0.10 \\
$M_{\rm bol}$, mag      &$-$10.41& 0.23      & $-$10.95& 0.25 \\
$DM$, mag               &   13.99& 0.24      & 14.52& 0.27 \\
$d$, kpc                &   6.3  & 0.7$^{c)}$  & 8.0& 1.0    \\
\hline
\multicolumn{5}{p{7.4cm}}{\footnotesize
$^{a)}$\,\cite{rauw05,rauw07} did not provide the uncertainties of $M_1$, $M_2$. We calculated their uncertainties, shown in the last column, using the uncertainties of $M_{1,2}\sin^3i$ from \cite{rauw05} and $i$ from \cite{rauw07}.

$^{b)}$\,The radii of the system components $R_1$, $R_2$ in our model are the radii of equal-volume spheres (see text).

$^{c)}$\,This distance estimate in our model is based on the same values of $R_V$ and $A_V$ as in \cite{rauw07} and may change given modern extinction data in the direction of WR\,20a (see text).
}
\end{tabular}
\end{table}
\renewcommand{\baselinestretch}{1.0}

Table~\ref{tab_abs_params} lists the absolute parameters of the system and its components found in our work and in \cite{rauw05,rauw07}. The radius of each component of the system in our model, $R_1$, $R_2$, is the radius of a sphere whose volume is equal to the volume of the star's body (the equipotential surface in the Roche model). As expected, in our model the Roche lobe filling factors, and hence the radii of the system components, turned out to be noticeably smaller (by about 25\%) than the values obtained in \cite{rauw07} with the standard Roche model. The reason for this is obvious: a model that does not take into account the component winds inevitably overestimates the radii of the stars in order to describe the increase in the width and depth of the light curve minima caused by the absorption in the winds (see Fig.~\ref{fig_lc_BVI}, top panel). Recall that we also run the model for the case where all parameters of the Roche model and winds were fixed and set equal to those from \cite{rauw05,rauw07}, except for the Roche lobe filling factors \mbox{$\mu_1=\mu_2$}. The resulting value $\mu_1=\mu_2=0.773$ was larger than the value $\mu_1=\mu_2=0.71$ in our final model. This is explained by the anticorrelation of $\mu$ and $\dot{M}$, discussed in Section~\ref{sec:results} . In our final model, the values of \mbox{$\dot{M}=1.24\times 10^{-5}\, M_\odot$\,yr$^{-1}$} are larger than $\dot{M}=0.85\times 10^{-5}\, M_\odot$\,yr$^{-1}$ adopted by \cite{rauw05}.

The mass-loss rate in \cite{rauw05} was determined from a spectral non-LTE model of the extended atmosphere based on the $\rm H\alpha$ line intensity, assuming a non-uniform wind and a volume filling factor $f=0.1$. In reality, the value of $f$ is rather uncertain. $\dot{M}$ in the spectral model scales approximately as $\sqrt{f}$. \cite{rauw05} note that in the case of a uniform wind, the value of $\dot{M}=2.5\times 10^{-5}\, M_\odot$\,yr$^{-1}$ would be required for the spectral model to reproduce the $\rm H\alpha$ and He\,II\,$\lambda\,4686$\,\AA~ line intensities. In our model, the optical opacity in the continuum is due to the electron scattering, which depends on the first power of density and, therefore, is independent of wind inhomogeneity. We conclude that the value $\dot{M}=1.24\times 10^{-5}\,M_\odot$\,yr$^{-1}$ obtained for the winds is a reasonable estimate.

The orbital inclination angle in our solution does not differ significantly from the value obtained in \cite{rauw07} while the temperature of the secondary component is somewhat higher (see Table~\ref{tab_found_params}). The luminosities of the system components and its total luminosity in our model are significantly lower than the values obtained in \cite{rauw07}. The main reason for this is the smaller values of the component radii in our model. As a result, the total luminosity of the system in our model is approximately 60\% of the total luminosity in the model of \cite{rauw07}. As a consequence, our estimate of the distance to the system also differs. Using the observed out-of-eclipse apparent stellar magnitude $V=13\fm416\pm 0\fm024$, the bolometric correction $BC=-3\fm 91$ and $A_V=5\fm93\pm 0\fm09$ from \cite{rauw07}, and the absolute bolometric stellar magnitude of the system in our model, we can calculate the modulus of the distance to the system $DM$ and the corresponding distance (see Table~\ref{tab_abs_params}).

Our distance estimate to the system, $d=6.3$\,kpc, is noticeably lower than the estimate $d=8.0$\,kpc in \cite{rauw07}. The authors of that paper determined spectrophotometric distances for a number of single O stars located in the young open cluster Westerlund~2; the average distance was $8.3 \pm 1.6$\,kpc. From this, they concluded that WR\,20a belongs to the cluster. Our distance estimate, at first glance, contradicts this conclusion. However, the authors of that paper themselves note that a number of factors can influence the distance estimate to the cluster. In addition, the distance to Westerlund~2 varies quite significantly in various literary sources. One of the reasons is that the cluster is located in the Carina spiral arm, in a region with strong extinction. Determining the distance to Westerlund~2 is far beyond the scope of the present paper. Therefore, we provide a brief (and not exhaustive) review of the literature, referring the interested reader to specific studies.

In earlier studies, distances were estimated from 2.5 to 8\,kpc (see the review by \citealp{rauw05}). \cite{russ03} estimated the kinematic distance to RCW\,49 (of which Westerlund~2 is a part) to be $d=4.7^{+0.6}_{-0.2}$\,kpc (but the author cautions that in the region of the Carina spiral arm there are deviations from the kinematic model of the Galactic rotation which he used). \cite{asc07} obtained a distance estimate $d=2.8$\,kpc by comparing photometry in $JHK_s$ filters with the evolutionary tracks of stars in the pre-MS stage. \cite{dame07} studied the CO emission of the giant molecular cloud [GCB88]\,8 (with a high probability associated with the cluster) and the absorption in the 21\,cm H\,I line in the direction of Westerlund~2. He found that the cluster may be located at the distant edge of the Carina spiral arm. Based on this, he determined the kinematic distance to the cluster \mbox{$d=6.0\pm 1.0$}\,kpc. By comparing the average velocities and velocity dispersion of gas in molecular clouds associated with RCW\,49, \cite{furuk09} determined the kinematic distance $d=5.4^{+1.1}_{-1.4}$\,kpc. \cite{rauw11} conducted a new spectroscopic study of 15 stars in Westerlund~2, including three eclipsing binaries (for which light curves were also analyzed). Distance estimates for the studied binaries ranged from 6\,kpc to more than 9\,kpc. The authors concluded that this result was consistent with their previous estimate of 8\,kpc \citep{rauw07}. \cite{vargas13} performed an extensive study of the cluster using Hubble Space Telescope data in several filters and ground-based photometric observations. They discovered 15 new O stars and determined individual $R_V$ and $A_V$ values for each star. The average values of these quantities were found to be $R_V=3.77\pm 0.09$ and $A_V=6\fm51\pm 0\fm38$. Recall that in \cite{rauw07} the values $R_V=3.1$ and $A_V=5\fm93$ were used for WR\,20a. An increase in $R_V$ and $A_V$ leads to a decrease in the distance to the cluster, which in \cite{vargas13} was estimated to be $d=4.2\pm 0.3$\,kpc. \cite{carraro13}, based on extensive observations in $UBVRI_cJHK_s$ filters, approximated the empirical dependence of $V$--$M_V$ on $E(B-V)$, and then estimated the value $R_V=3.88\pm 0.18$, but obtained a distance \mbox{$d=2.85\pm 0.43$}\,kpc that is significantly smaller than in \cite{vargas13}. \cite{zeid15} performed a multi-band study of the cluster with high spatial resolution on the Hubble Space Telescope and obtained the average value $R_V=3.95\pm 0.14$. The authors note that the absorption in the cluster region is non-uniform. \cite{lopes24} refined the spectral classification of several O stars in Westerlund~2, classifying two of them for the first time as belonging to the spectral type O2\,V, and re-estimated the temperatures of the stars studied. Comparing the observed positions of five stars on the Hertzsprung-Russell diagram with the positions they should occupy based on the theory of massive stars evolution, he estimated the distance to the cluster $d\simeq 5$\,kpc. The distances to the same stars from the Gaia\,EDR3 catalog, determined by both the geometric (rgeo) and photogeometric (rpgeo) methods, are $5.3\pm 1.5$\,kpc.

We also used Table {\tt gedr3dist.main} from the Gaia\;EDR3 catalog (\citealp{gaia16}; \citealp{gaia_edr3_21}), whose distances are calculated in \cite{bailer21}, to extract the distances to the O stars listed in Table~1 in \cite{rauw07}. Three stars (Cl* Westerlund\,2 MSP\,157, MSP\,188, MSP\,203) are absent from the catalog. For the remaining nine stars, the average distances determined by the two methods are ${\rm rgeo}=4.48\pm 0.21$\,kpc and ${\rm rpgeo}=4.33\pm 0.21$\,kpc.

Note that, according to the Gaia\,EDR3 catalog, the distance to WR\,20a itself, determined by the same methods, was $d=4.5^{+0.4}_{-0.3}$\,kpc. The distance of 6.3\,kpc, obtained in our model and given in Table~\ref{tab_abs_params}, is based on the values of the bolometric correction of WR\,20a and the quantities $R_V=3.1$, $A_V=5\fm93\pm 0\fm09$ from \cite{rauw07}. As follows from our brief review above, the value of $R_V$ for WR\,20a and Westerlund~2 can be significantly larger than the standard value of 3.1 (up to the value of 4). This means that $A_V$ will also increase, and the distance will decrease. It is easy to verify that in order for the distance corresponding to our solution to become equal to 4.5\,kpc, a relatively small increase in $R_V$ to a value of 3.48 is necessary.

The distances to Westerlund~2 and WR\,20a derived from Gaia data should be treated with caution. Gaia distances are quite accurate for relatively nearby objects. For objects located at distances of several kiloparsecs, the parallaxes are very small, and Gaia estimates may contain large random and systematic errors. An example of such systematic errors is the recent work of \cite{majaess25}, which compares the Gaia distance to the Pismis\,19 cluster, approximately 3.5\,kpc according to DR3 data, with the distance to this cluster of approximately 2.5\,kpc, obtained by conventional methods. Nevertheless, a number of recent conventional studies, listed above, suggest that $R_V$ in the direction of Westerlund~2 is significantly larger than the canonical value of 3.1, leading to distances of the order of 4\,kpc. As shown above, using modern estimates of $R_V$ in our model yields a distance to WR\,20a that is consistent with recent distance estimates to Westerlund~2. Thus, our results are consistent with the hypothesis that WR\,20a is a member of a cluster.

\section{Conclusions}\label{sec:concl}

The proposed model of a binary system with two colliding winds successfully describes the observed light curves of WR\,20a. Taking into account the absorption of radiation from the binary components in the winds and cooling layers leads to smaller values of the component radii compared to the values obtained by \cite{bonanos04} and \cite{rauw07} in the standard Roche model. This demonstrates the importance of taking into account the presence of stellar winds when interpreting light curves of such binary systems. The model was able to adequately reproduce the asymmetry of the observational light curves with respect to the conjunction phases, in contrast to the standard Roche model, in which the configuration of the system in a circular orbit (and, consequently, the model light curve) is symmetric with respect to these phases. A comparison of the $BVI$ light curves obtained in 2004–2005, with the ASAS-SN (2016–2018) and TESS (2019–2023) light curves, as well as an examination of the non-averaged TESS light curve, shows that the system exhibits irregular variability, likely due to non-stationary processes occurring in the wind-wind collision zone. Our steady-state model describes the average state of the collision zone, and therefore the model light curves should reproduce the average observational curves. This is clearly demonstrated by Fig.~\ref{fig_lc_TESS_ASAS}, where the model light curve reproduces well the average observational curve of TESS, despite the fact that the model was optimized using different data (in the $BVI$ filters).

The luminosity of the system in our model is approximately 40\% smaller than that in \cite{rauw07}. The distance to the system is approximately 20\% smaller, which, at first glance, contradicts the hypothesis that WR\,20a is a member of the young open cluster Westrlund~2. However, taking into account recent studies of interstellar extinction in the direction of this cluster, the distance to WR\,20a obtained in our model is consistent with the hypothesis that WR\,20a is a member of the cluster.

\begin{acknowledgments}

This paper includes data collected by the TESS mission, which are publicly available from the Mikulski Archive for Space Telescopes (MAST). Funding for the TESS mission is provided by NASA’s Science Mission directorate. We acknowledge the use of the data from ``All-Sky Automated Survey for Supernovae'' (ASAS-SN), which is publicly available from the ASAS-SN Photometry Database. This work has made use of data from the European Space Agency (ESA) mission Gaia (\url{https://www.cosmos.esa.int/gaia}), processed by the Gaia Вata Processing and Analysis Consortium (DPAC, \url{https://www.cosmos.esa.int/web/gaia/dpac/consortium}). Funding for the DPAC has been provided by national institutions, in particular the institutions participating in the Gaia Multilateral Agreement.

\end{acknowledgments}

\section*{Funding}

The work of I.I.A. (development of the algorithm for accounting for two stellar winds in a binary system, the MCMC method, and analysis of observational data) was supported by grant No. 23-12-00092 of the Russian Science Foundation. The work of E.A.A. was supported by the Interdisciplinary Scientific and Educational School of M.V. Lomonosov Moscow State University ``Fundamental and Applied Space Research''.

\section*{Conflict of interest}

The authors of this work declare that they have no conflicts of interest.

\bibliographystyle{aspb3}
\bibliography{Antokhin}

\end{document}